\documentclass[preprint,authoryear,12pt]{elsarticle}

\usepackage{amssymb}
\usepackage{graphicx}
\journal{New Astronomy}
\def\astrobj#1{#1}
\begin{document}

\begin{frontmatter}



%\title{Absolute Dimensions of the Contact Binary BO CVn}
\title{Spectroscopic and Photometric Study of the Contact Binary BO CVn}


\author[1,2]{S. Zola}
\author[3,4]{R.H. Nelson}
\author[5]{H.V. \c{S}enavc{\i}}
\author[1]{T. Szymanski}
\author[1]{A. Ku\'zmicz}
\author[2]{M.~Winiarski}
\author[1]{D. Jableka}

\address[1]{Astronomical Observatory, Jagiellonian University, ul. Orla 171, 
30-244 Krakow, Poland}
\address[2]{Mt. Suhora Observatory, Pedagogical University, ul. Podchorazych 2,
30-084 Krakow, Poland}
\address[3]{1393 Garvin Street, Prince George, BC, Canada, V2M 3Z1}
\address[4]{Guest investigator, Dominion Astrophysical Observatory, Herzberg Institute of Astrophysics,
National Research Council of Canada}
\address[5]{Ankara University, Faculty of Science, Dept. of Astronomy and Space Sciences, TR-06100 Tando\u{g}an-Ankara, TURKEY}

\begin{abstract}

We present the results of the study of the contact binary system
\astrobj{BO CVn}. 
We have obtained physical parameters of the components based on 
combined analysis of new, multi-color light curves and spectroscopic
mass ratio.  This is the first time the latter has been determined for 
this object.  
We derived
the contact configuration for the system with a very high filling factor of 
about 88 percent. We were able to reproduce the observed light curve, namely
the flat bottom of the secondary minimum, only if a third light has been added 
into the list of free parameters. The resulting third light contribution 
is significant, about 20-24 percent, while the absolute parameters of 
components are: M$_{1}$=1.16, M$_{2}$=0.39, R$_{1}$=1.62 and R$_{2}$=1.00 
(in solar units). 
 
The O-C diagram shows an upward parabola which, under the conservative 
mass transfer assumption, would correspond to a mass transfer rate of 
$dM/dt$  =  $6.3\times10^{\rm -8} M_{\odot}/yr$, matter being 
transferred from the less massive component to the more massive one. 
No cyclic, short-period variations have been found in the O-C diagram
(but longer-term variations remain a possibility).

\end{abstract}


\end{frontmatter}

% \linenumbers

\section{Introduction}
\label{intro}
The light variability of BO CVn (9.9~mag) was first discovered by \citet{oja89}, who 
published the first
U, B and V light curves of the system with an initial period of 0.517460 days and also classified
BO CVn as a W UMa type eclipsing variable. With the help of the Hydrogen 
($H_{\gamma}$ and $H_{\delta}$)
and CaII K lines, taken using the objective prism plates, \citet{oja89} determined 
the spectral type of the system as F0.

The first light curve and period analysis of the system were performed by \citet{albayrak01}, using
the normal points derived from the B and V light curves of the system obtained with a SSP-5A photometer
attached to the 30 cm Maksutov telescope at the Ankara University Observatory. 
They used the revised version of the Wilson-Devinney code \citep{wilson71} for the study. 
Due to the lack of spectroscopically determined mass  ratio, \citet{albayrak01} performed 
a “so-called q search” and determined the best value of the mass ratio as q = 0.205. 
As the results they  showed that BO CVn is in the contact configuration
 with a filling factor of $f$ = 0.18 and a temperature difference of ${\Delta}T$ = 90 K. 
From the period (O-C) analysis, assuming that the period change is due to mass transfer, 
they determined the period increase of 0.037 s/yr, that under the conservative case, 
corresponds to mass transfer rate of 1.57 x 10$^{-10}$ $M_{\odot}$/yr (matter being
transferred from the less massive to the more massive component). 

Another light curve and period analysis of BO CVn were carried out by \citet{qian06}. Using 
the B, V and R light curves obtained with PI1024 TKB CCD attached to the 1.0 m reflecting 
telescope at the Yunnan Observatory of China, they analyzed their original data points 
with the 2003 version of the Wilson-Devinney code.
Similarly to the procedure done by \citet{albayrak01}, \citet{qian06} obtained a solution 
for the photometric mass ratio $q$ = 0.204, very close indeed to that  obtained by 
\citet{albayrak01}. 
They also determined the value of $M_{1}$ as 1.70 $\pm$ 0.21 $M_{\odot}$ , 
using the statistical relation between $M_{1}$ and q for hotter contact binaries 
and estimated the value of $M_{2}$ as 0.35 $M_{\odot}$.

\begin{table}
\caption{The log of spectroscopic observations
}
\begin{center}
\begin{scriptsize}
\begin{tabular}{cccrcll}
\hline
 DAO          & Mid Time       & Exposure   & S/N& Phase at & ~~~V$_1$  & ~~~V$_2$  \\
 Image \#     &(HJD - 2400000) & Time (sec) & &Mid-Exposure & (km/s) & (km/s) \\
\hline
4801 & 54569.0244 & 2448 &  75 & 0.255 &   -56.7$\pm$7.9 & ~245.9$\pm$11.2\\
4851 & 54571.8556 & 3600 &  95 &0.726 &    ~85.6$\pm$10.7&  -226.0$\pm$5.0\\
4886 & 54574.6943 & 3600 & 109 &0.212 &    -60.2$\pm$9.3 & ~227.4$\pm$12.9\\
4888 & 54574.7383 & 3600 & 115 &0.297 &    -52.2$\pm$10.2& ~241.2$\pm$16.8\\
5298 & 54925.8028 & 3600 & 116 &0.731 &    ~89.5$\pm$5.4 &  -220.1$\pm$4.7\\
5339 & 54926.8407 & 1800 &  82 &0.736 &    ~92.8$\pm$3.4 &  -218.6$\pm$8.1\\
5387 & 54928.7197 & 3600 &  99 &0.368 &    -25.8$\pm$4.4 & ~198.8$\pm$11.2\\
5432 & 54929.9678 & 3103 &  59 &0.780 &    ~95.2$\pm$3.2 &  -215.9$\pm$4.2\\
2650 & 55673.8127 & 3600 & 111 &0.264 &    -56.1$\pm$3.9 & ~238.5$\pm$8.8\\
2740 & 55677.9635 & 3600 &  88 &0.285 &    -48.3$\pm$6.1 & ~243.6$\pm$4.9\\
\hline
\end{tabular}
%\end{scriptsize}
%\begin{center}
\begin{center}
V$_1$, V$_2$ are radial velocities of primary and secondary components, respectively 
\end{center}
\end{scriptsize}
\end{center}
\label{SpecObsElm}
\end{table}

In this study, Johnson V, R and Str\"{o}mgren v, b, y light curves were analyzed making use of
the system spectroscopic mass ratio value, determined for the first time, in order to obtain
accurate absolute parameters of the components of the BO CVn eclipsing binary. 
In addition, an up-to-date analysis of the system period behavior was performed. 

\section{Observations and Data Reduction}

\subsection{Photometry}
\label{photometry}
\astrobj{BO CVn} was observed photometrically with the aim to obtain as accurate as possible, 
multicolor light curve. The observations in the Str\"{o}mgren v and b filters have
been gathered at  Mt. Suhora Observatory of the Pedagogical University using the 
60~cm telescope and an Apogee Alta U47 CCD, mounted at the primary focus of the telescope.
The 50~cm telescope at the Astronomical Observatory of the Jagiellonian University in 
Krakow, Poland, equipped with an Andor DZ 936-BV CCD, installed at the Cassegrain focus, was  
used to take additional data in the  Str\"{o}mgren y and wide band V and R (Bessell) filters. 
 We used the same comparison star (\astrobj{GSC 3030 1129}, 8.5~mag, K0) at both sites while
\astrobj{GSC 3030 1123} served as the check star (10.7~mag, B-V=0.74). The light curve in Str\"{o}mgren
v and b filters has been gathered during four nights: March 27th, 29th, April 10th and 17th,
2011.
Observations in Krakow (in Str\"{o}mgren y and wide band V and R filters) were taken on March
8th, 28th, 29th 30th and April 3rd, 2011.  
 Calibration of frames has been performed in a standard way:  using the IRAF package 
all scientific images have been calibrated for bias, dark and flatfield which have
been taken each night.
 The magnitude 
differences between variable, comparison and check stars were derived by means of the aperture
photometry making use of the CMunipack program 
({\it{http://c-munipack.sourceforge.net/, March 30th, 2012}}),  
which is an interface for the DAOPHOT package.
Due to the large color difference between our target and chosen comparison star we accounted
for both differencial and color terms of atmospheric extinction.   
The observations have been phased using the following linear ephemeris, taken from the Kreiner's
on-line catalogue \cite{kreiner04} ({\it{http://www.as.up.krakow.pl/minicalc/CVNBO.HTM, August 29th, 2011}}):

\begin{equation}
HJD~(Min~I) = 2452500.0763 + 0^{\rm d}.5174629 \times E.
\end{equation}

The data in V and R filters taken at different nights were combined smoothly. However, those
gathered in the Str\"{o}mgren filters showed a small but noticeable shift for one night
(March 27/28, 2011).
To make them fit, we have moved the outlying night points by the values calculated
from the overlapping part of the light curve.
The final light curve consists of 1633, 1624, 1616, 933 and 994
individual points in v, b, y, V and R filters, respectively.
The new multicolor light curve shows no or negligible difference in the maxima heights 
and the flat-bottom secondary minimum is clearly visible. 
       
\begin{figure}
\center
\includegraphics*[width=6.0cm,height=14.0cm,scale=1.0,angle=270]{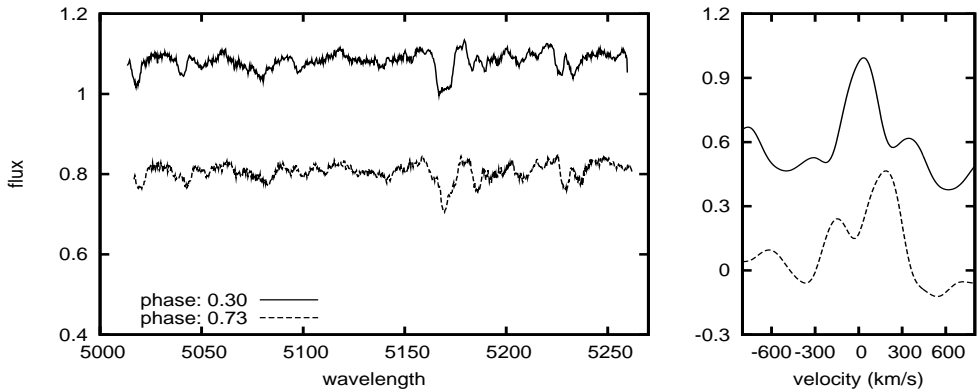}
 \caption{The spectra of BO CVn (left panel) taken at both quadratures and 
corresponding broadening functions (right panel).}
\label{SpecBF}
\end{figure}

\subsection{Spectroscopy}
\label{spectra}
\vskip 0.4cm

During three nights in 2008, four in 2009, and two in 2011 (all in April),
a total of 10 spectra (at 10 $\rm{\AA}$/mm reciprocal dispersion, resolving
power = 10,000) were taken by R.H. Nelson at the Dominion Astrophysical Observatory (DAO) in Victoria,
British Columbia, Canada; then the Rucinski's broadening functions \citep{rucinski04} were used
to obtain radial velocity (RV) curves (see \citealt{nelson06} and \citealt{nelson10b} for details).
The spectral range was approximately 5000-5260 $\rm{\AA}$. 
A log of the DAO observations and
RV results are presented in Table \ref{SpecObsElm}, while two spectra taken at phases: 0.30 and
0.73 and broadening functions at these phases are shown in Fig. \ref{SpecBF}.
The fit of the sinusoidal curves gave the following orbital parameters: 
V$_{\gamma}$=15.4$\pm{1.8}$, K$_{1}$=72.5$\pm{2.3}$ and K$_{1}$=233.6$\pm{3.0}$ km/s.
The spectroscopic mass ratio value 
q$_{spec}$=0.31$\pm{0.01}$ (M$_{2}$/M$_{1}$) is significantly higher than that derived by  
\citet{albayrak01} and  \citet{qian06} from the light curve modelling alone. 
This is not surprising, as \cite{terrellwilson2005} showed numerically that the 
photometric mass ratio is unreliable for contact binaries exhibiting partial eclipses.
   
\section{The Light Curve Modeling}

 All previous \astrobj{BO CVn} light curves modelling  gave a contact configuration.
However, without knowing the spectroscopic mass ratio value, any attempt to derive the
parameters of components may lead to spurious values. 
This situation has improved with the q$_{spec}$ mass ratio determined and we performed
the analysis of the new, multicolor light curve. In order to speed up computations
mean points in each filter have been calculated: 141 in v and b filters, 139, 134 and 136
in the y, V and R filters, respectively. Furthermore, the magnitudes of the mean points
have been recalculated to the flux units and normalized to unity at the first maximum.

\begin{figure}
\includegraphics*[width=8.0cm,height=13.8cm,scale=1.0,angle=270]{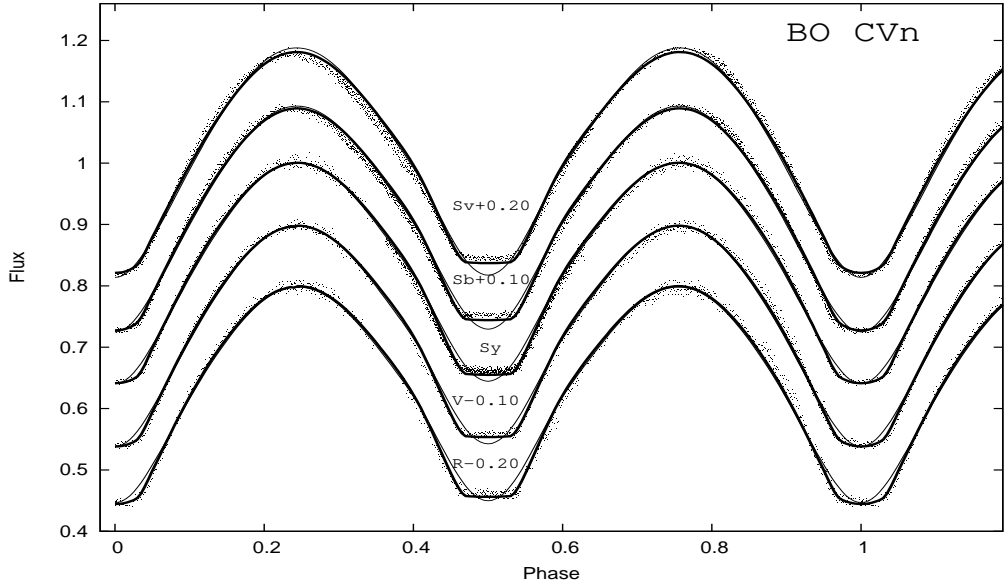}
 \caption{Comparison between theoretical and observed light curves of BO~CVn
  (vbyVR filters, models with and without l$_{3}$). 
  Dots represent individual observations, theoretical light curves are shown as 
 continuous lines: thin for l$_{3}$=0 and thick for l$_{3}$ adjusted.}
\label{PhotModel1}
\end{figure}

 We used the Wilson-Devinney code (\citealt{wilson79, wilson93}, hereafter W-D),  
appended with the Monte Carlo search method. 
This code has been already applied to solve light curves of dozens of
systems (see \citealt{zola10} and references therein). 
Instead of the original differential correction method it deploys
the Monte Carlo as the search algorithm, greatly improving the chances of finding the
global solution in the searched range of parameters. It does not require to
assume any initial system configuration - it is found as the result of computations.
To avoid the problem of arbitrary assumed weights to a few radial velocities measurements
and photometric data (a thousand or more) we decided to perform the computations in
an iterative way. 
The photometric light curve has been solved separately, then, after
solution has converged, the results from this step serve for the mass ratio determination
using the original W-D code and the radial velocity measurements. Only parameters relevant 
to the orbit are adjusted in this step. The recomputed mass ratio value is used for
another run, when the best solution of the photometric light curve is found and such
a procedure was repeated until the $q$ corrections become smaller than the mass ratio
error.

\begin{figure}
\includegraphics*[width=7.5cm,height=12.5cm,scale=1.0,angle=270]{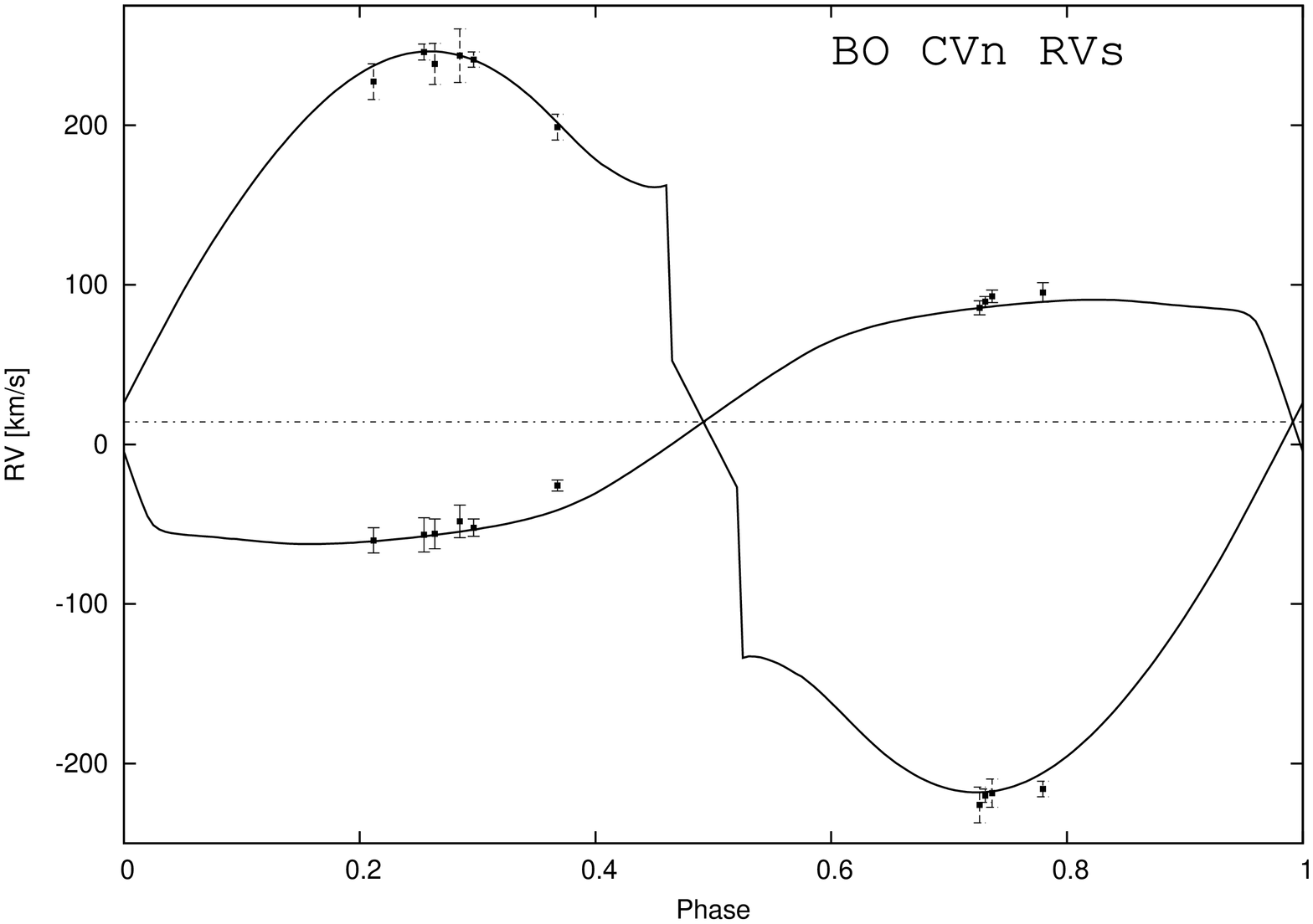}
 \caption{Comparison between theoretical and observed radial velocity curves of BO~CVn.
  Squares represent individual observations, theoretical radial velocity curves are shown by 
  continuous lines.}
\label{QspecRVModel}
\end{figure}

  Several parameters have been fixed or calculated in our modelling: the temperature of
the primary component was set according to its F0 spectral type at 6980~K \citep{harmanec88},
the values for albedo and gravity darkening coefficients have been chosen at their
theoretical values while the darkening coefficients  adopted from the \citet{diaz95}
tables as functions of the temperature and wavelength. Furthermore, we adopted the values 
for albedo and gravity darkening to agree with theoretical values: 0.50 and 0.32, respectively,
appropriate for a convective stellar envelope.
The following parameters have been
adjusted: inclination (70-90 degrees), temperature of the secondary (6000-7500~K), components 
potentials (2-3.8) and luminosity of the primary (4-13.5). The ranges
of fitted parameters have been listed in the parentheses. 

\begin{table}
%\begin{center}
%\begin{flushleft}
\caption{Results derived from the light curve modelling. The adjusted parameters 
are: inclination (i), temperature of the secondary component (T$_{\rm 2}$),
potentials ($\Omega$), the primary component luminosity (L$_{1}$) and
third light ($l_{3}$). Given uncertainties are those derived from the fit 
at the 90\% confidence level. The temperature of the primary (T$_{\rm 1}$) and
the system mass ratio (q$_{corr}$) have been kept fixed at this step. With 
the IPB control parameter set to 0, luminosity of the secondary (L$_{2}$) was
calculated by the W-D code. } 
\begin{center}
\begin{small}
\begin{tabular}{lr}
\hline
parameter            &    value             \\
\hline
filling factor       &  88 \%                \\
phase shift          &  0.0002$\pm$0.0001     \\
$i$ (deg)            &  89.4$\pm$0.2         \\
$T_{\rm 1}({\rm K})$ &  * 6980                \\
$T_{\rm 2}({\rm K})$ & 7191 $\pm$ 10          \\
$\Omega_{\rm 1} $ = $\Omega_{\rm 2}$ & 2.3796 $\pm$0.0002   \\
$q_{corr}$ ($M_{\rm 2}$/$M_{\rm 1})$ & * 0.340                \\
\hline
$L_{1}~(v)$           & 5.853 $\pm$0.018      \\
$L_{1}~(b)$           & 6.055 $\pm$0.018      \\
$L_{1}~(y)$           & 6.287 $\pm$0.020      \\
$L_{1}~(V)$           & 6.262 $\pm$0.019      \\
$L_{1}~(R)$           & 6.427 $\pm$0.022      \\
$L_{2}~(v)$           & ** 2.820              \\
$L_{2}~(b)$           & ** 2.874              \\
$L_{2}~(y)$           & ** 2.935              \\
$L_{2}~(V)$           & ** 2.922              \\
$L_{2}~(R)$           & ** 2.953              \\
$l_{3}~(v)$           &   0.2400 $\pm$0.0014   \\
$l_{3}~(b)$           &   0.2276 $\pm$0.0015   \\
$l_{3}~(y)$           &   0.2164 $\pm$0.0017   \\
$l_{3}~(V)$           &   0.2168 $\pm$0.0017   \\
$l_{3}~(R)$           &   0.2035 $\pm$0.0021   \\
\hline
\end{tabular}
\end{small}
%\end{flushleft}
\begin{center}
\begin{small}
*-not adjusted, **-computed
\end{small}
\end{center}
\end{center}
\label{ResTab}
\end{table}

 The spectroscopic mass ratio value has been estimated from the spectroscopic observations 
to be q$_{spec}$=0.31 and this value was kept fixed in the first run. During subsequent runs this
value has been modified to q$_{corr}$=0.34 and further correction have been smaller than the
estimated error of the fit. 
The modified mass ratio value being the result of subsequent fits to the RV curves using the
W-D code (only orbital parameters have been adjusted) is of somewhat higher value than 
q$_{spec}$ but the recomputed q accounts for proximity effects.  
It soon turned out that for the spectroscopic mass ratio value it was not possible
to derive a model showing the flat bottom secondary minimum.  
We were able to get a good description
of the observed light curve only if another free parameter is added - the third light (l$_3$). 
Only then did the model fit the flat bottom shape of the secondary minimum.

\begin{figure}
\center
\includegraphics[width=10.0cm,height=9.0cm,scale=1.2,angle=0]{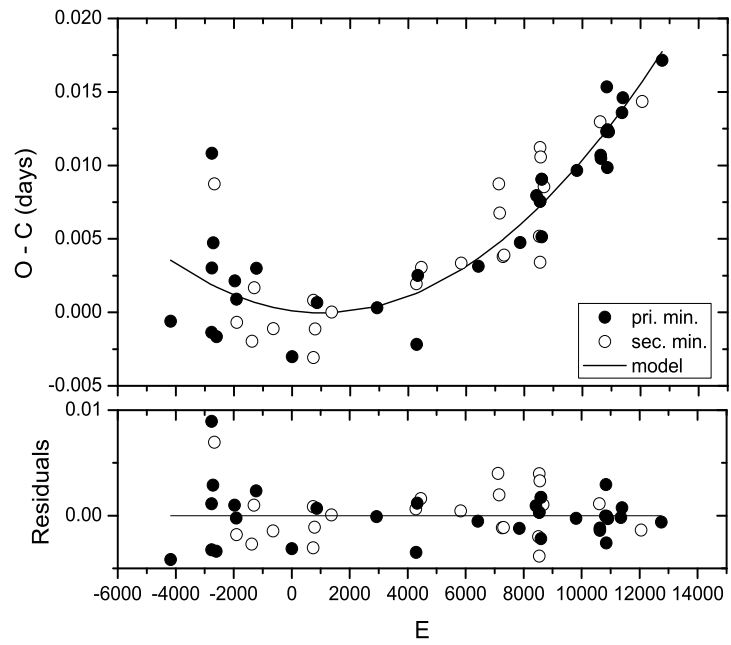}
 \caption{The O-C diagram of BO CVn with the secular variation approximation and 
the residuals from the model.}
\label{oc}
\end{figure}

The final results are presented in Tab. \ref{ResTab}  and the theoretical light curve 
along with the observed one is shown in Fig. \ref{PhotModel1}. In Fig. \ref{QspecRVModel}
the theoretical radial velocities (lines) along with measured ones (squares) are shown.
As it can be seen, 
the model without a third light (thin lines) gives a very poor fit in both minima. 
When a third light parameter has been adjusted, a much better fit was obtained 
(thick lines) and the shape of the secondary minimum resembles that observed.  
In Fig. \ref{QspecRVModel} the theoretical and observed radial velocities calculated 
for the best fit parameters are shown.

\section{The orbital period variation}
\label{orb}
The O-C analysis of BO CVn was performed in order to investigate the orbital period variation 
behavior of the system. All available minima times, which cover 24 years (between 1987 and 2011), 
were used during the analysis and listed in Table \ref{OCTab} together with the minimum time 
obtained in this study.

\begin{table}
\caption{Available minima times for BO CVn}
\center
\scriptsize
\begin{tabular}{cccccccccc}
\hline
 HJD Min & Type & Meth & O - C & Ref. & HJD Min & Type & Meth & O - C            &  Ref.             \\
 (2400000+)& & & & & & & & & \\
\hline
46895.4550 & 1 & pe & -0.00060 & 1 & 52757.5142 & 2 & pe & 0.00675 & 13 \\
47624.5540 & 1 & pe & -0.00136 & 2 & 52813.3986 & 2 & pe & 0.00379 & 13 \\
47624.5550 & 1 & pe & -0.00136 & 2 & 52835.3914 & 1 & ccd & 0.00390 & 14 \\
47626.6300 & 1 & pe & 0.00301 & 2 & 53119.2196 & 2 & ccd & 0.00475 & 15 \\
47626.6380 & 1 & pe & 0.01083 & 2 & 53413.3989 & 1 & ccd & 0.00795 & 16 \\
47651.4684 & 1 & pe & 0.00474 & 3 & 53413.3990 & 1 & ccd & 0.00795 & 16 \\
47651.4690 & 1 & pe & 0.00474 & 3 & 53413.3991 & 1 & ccd & 0.00795 & 16 \\
47673.4643 & 2 & pe & 0.00874 & 4 & 53461.5214 & 1 & ccd & 0.00518 & 17 \\
47709.4180 & 1 & pe & -0.00166 & 4 & 53475.2303 & 2 & ccd & 0.00341 & 16 \\
48036.4588 & 1 & pe & 0.00215 & 4 & 53475.2321 & 2 & ccd & 0.00341 & 16 \\
48065.4322 & 1 & pe & 0.00091 & 4 & 53475.2376 & 2 & ccd & 0.01122 & 16 \\
48071.3841 & 2 & pe & -0.00067 & 4 & 53476.2681 & 2 & ccd & 0.00755 & 16 \\
48341.4974 & 2 & pe & -0.00195 & 4 & 53476.2691 & 2 & ccd & 0.00755 & 16 \\
48383.4156 & 2 & pe & 0.00169 & 4 & 53476.2692 & 2 & ccd & 0.00755 & 16 \\
48419.3795 & 1 & pe & 0.00300 & 4 & 53485.3277 & 1 & ccd & 0.01058 & 14 \\
48724.4164 & 2 & pe & -0.00110 & 5 & 53504.4707 & 1 & ccd & 0.00515 & 14 \\
48724.4187 & 2 & pe & -0.00110 & 5 & 53504.4714 & 1 & ccd & 0.00906 & 14 \\
49056.3672 & 1 & pe & -0.00301 & 2 & 53504.4728 & 1 & ccd & 0.00906 & 14 \\
49056.3682 & 1 & pe & -0.00301 & 2 & 53504.4735 & 1 & ccd & 0.00906 & 14 \\
49439.5470 & 2 & pe & -0.00308 & 6 & 53545.3509 & 1 & pe & 0.00856 & 18 \\
49439.5490 & 2 & pe & 0.00083 & 6 & 54131.8950 & 2 & ccd & 0.00966 & 19 \\
49465.4221 & 2 & pe & -0.00112 & 6 & 54547.1619 & 1 & ccd & 0.01296 & 20 \\
49465.4223 & 2 & pe & -0.00112 & 6 & 54554.4039 & 1 & ccd & 0.01070 & 21 \\
49503.4554 & 1 & pe & 0.00066 & 6 & 54557.7673 & 2 & ccd & 0.01048 & 22 \\
49503.4555 & 1 & pe & 0.00066 & 6 & 54656.3429 & 1 & ccd & 0.01232 & 23 \\
49763.4798 & 2 & pe & 0.00003 & 7 & 54665.4026 & 2 & ccd & 0.01535 & 23 \\
49763.4803 & 2 & pe & 0.00003 & 7 & 54671.3491 & 1 & ccd & 0.00986 & 23 \\
50570.4605 & 1 & pe & 0.00031 & 8 & 54678.3377 & 2 & ccd & 0.01242 & 23 \\
51271.3630 & 2 & pe & 0.00195 & 9 & 54699.2940 & 1 & ccd & 0.01229 & 23 \\
51295.4252 & 1 & pe & 0.00252 & 9 & 54933.4463 & 2 & ccd & 0.01360 & 24 \\
51362.4360 & 2 & pe & 0.00306 & 10 & 54950.7803 & 1 & ccd & 0.01460 & 25 \\
52070.3231 & 2 & pe & 0.00334 & 11 & 55294.3759 & 1 & ccd & 0.01435 & 26 \\
52375.3677 & 1 & pe & 0.00315 & 11 & 55651.4263 & 1 & ccd & 0.01715 & 27 \\
52740.4401 & 2 & pe & 0.00874 & 12 \\          

\hline
\end{tabular}
\label{OCTab}
References: 1. \citet{oja89}, 2. \citet{hubscher93}, 3. \citet{hubscher91}, 4. \citet{albayrak00}, 5. \citet{hubscher92}, 6. \citet{hubscher94}, 7. \citet{agerer96}, 8. \citet{agerer99}, 9. \citet{agerer00}, 10. \citet{agerer01}, 11. \citet{albayrak02}, 12. \citet{muyes03}, 13. \citet{selam03}, 14. \citet{brat07}, 15. \citet{nagai05}, 16. \citet{qian06}, 17. \citet{hubscher05}, 18. \citet{albayrak05}, 19. \citet{nelson08}, 20. \citet{nagai09}, 21. \citet{brat08}, 22. \citet{nelson09}, 23. \citet{yilmaz09}, 24. \citet{brat09}, 25. \citet{nelson10a}, 26. \citet{hubscher11}, 27. This study
\end{table}

All the previous O-C analysis of the contact binary BO CVn (see Section \ref{intro}) 
indicated a period increase of the system. Therefore, as the first attempt, 
we tried to determine the best fit to this secular variation using an 
IDL\footnote[1]{http://ittvis.com/ProductServices/IDL.aspx} routine, relying on the 
Levenberg-Marquardt algorithm. The resulting quadratic light elements  (including the
errors) are given below:

\begin{equation}
%HJD~(Min~I) = 2449056.3702(4) + 0^{\rm d}.5174605(2) \times E + 1.3063(2) \times 10^{\rm -10} \times E^{\rm 2}.
Min~I_{HJD} = 2449056.3702(4) + 0^{\rm d}.5174605(2) \times E + 1.3063(2) \times 10^{\rm -10} \times E^{\rm 2}.
\end{equation}

The O-C diagram with the model are shown in Fig. \ref{oc}, including the residuals from 
the fit. As it can be seen from Fig. \ref{oc}, the residuals from the quadratic fit show no 
any short-period cyclic variation that can be attributed to either the Light-Time Effect 
or some magnetic activity of the system. However, the possibility for long-term variation 
due to a light time effect remains a possibility.

The parameters derived from the O-C analysis, with the help of the absolute dimensions 
(masses of both components) determined from the photometric analysis are given in Table \ref{OCTab2}.

\section{Results and Discussion}

 We gathered new, accurate, multicolor light curves of the eclipsing binary system BO CVn.
%Our observations revealed the flat bottom shape of the secondary minimum for the first time.
Ten spectra taken around the quadratures make it possible to derive the spectroscopic mass
ratio of the system for the first time. 

\begin{table}
\center
\caption{Results derived from the O-C analysis }
\begin{tabular}{lcc}
\hline
parameter       &   value      &   error             \\
\hline
$dP/dE$ (days/cycle)   &   $2.6\times10^{\rm -10}$         &  $3\times10^{\rm -11}$         \\
$dP/dE$ (days/year)   &   $1.8\times10^{\rm -7}$         &  $2\times10^{\rm -8}$         \\
$dM/dt$ ($M_{\odot}$/year)   &   $6.3\times10^{\rm -8}$     & $2\times10^{\rm -9}$         \\
\hline
\end{tabular}
\label{OCTab2}
\end{table}

 We made light curve modelling of the new light curves with the W-D code and making use of 
the spectroscopic mass ratio. 
It turned out that it was not possible to obtain a good fit without an additional free
parameter: the third light. Only when the third light was allowed to be adjusted,
we could achieve the observed shape of the secondary minimum. The theoretical
light curves fit observations reasonably well, with very small departures at the beginning of
the descending branch of the secondary and primary minima. 
As mentioned in Section \ref{photometry}, we found a problem in combining the data from 
a single night, covering these phases. 
Therefore, we can only speculate that despite of all our attempts, not all atmospheric 
effects have been reduced.  There could be also another reason for this
asymmetry in the light curve: some  magnetic activity  resulting in spot(s)
appearing on the surface of one or both stars. Large, cool polar spots have been reported 
on the surface of \astrobj{31 Com} despite its thin convective layer \citep{strassmeier10}. 

\begin{figure}
\center
\includegraphics[width=7.0cm,height=10.0cm,scale=1.0,angle=270]{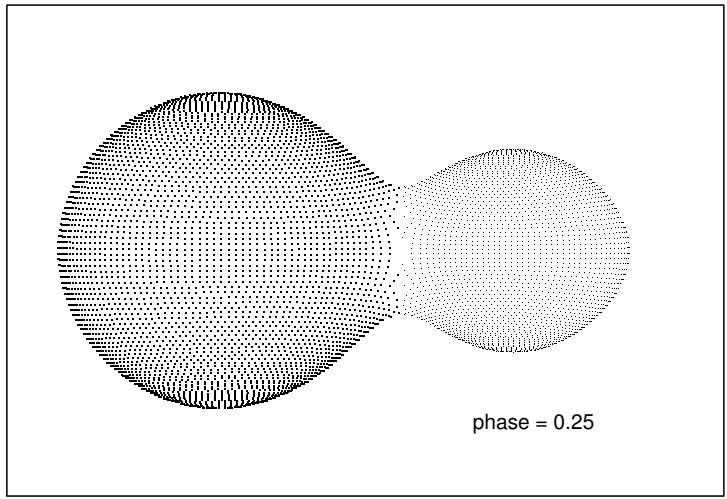}
\caption{A 3D model of BO CVn resulting from our solution shown at phase 0.25.}
\label{3dmodel}
\end{figure}

 Our light curve modelling results indicate that \astrobj{BO CVn} is a deep contact 
system with a very high filling factor of 88 percent. 
A view of the system configuration is shown in Fig. \ref{3dmodel}. 
The secondary, a smaller star, is about 200~K hotter than the
primary one. We derived a high system inclination, close to 90 degrees. 
The contribution of the third light is significant, reaching about 24 percent in 
the Str\"{o}mgren v filter.

Spectra exposure times compared to the orbital period were long and that 
must cause the phase smearing that reduces the  amplitude of the measured 
radial velocity variations. 
We recomputed the amplitudes accounting for this effect and derived the 
following values: K$_{1}$=74.9$\pm{2.6}$ and K$_{2}$=241.8$\pm{2.6}$ km/s.   
Combining the results from spectroscopy and those 
derived from light curve modelling, we calculated the absolute system parameters, 
listed in Table~\ref{TabAbsParam}. 
While parameters of the primary place it at or near the Main Sequence, the secondary 
is much oversized for its mass. 
The errors of parameters are given for the 90\% confidence level while radius is that
in the side direction. 
It is worth noting here that our values of physical parameters, 
which are based on the spectroscopic observations and those derived by \citet{qian06} based on
the ``photometric'' mass ratio, differ significantly. As mentioned in Sect. \ref{spectra} 
an application of the ``q-search'' to solve the light curves of partial eclipsed, contact
binaries most likely will result in unreliable results.

Since the O'Connell effect is not clearly visible, we decided to present
a no-spot solution as the final one. However, if indeed a spot presence is
responsible for the small discrepancy between the theoretical and observed light curves, we
eventually checked how this would influence the physical parameters and their uncertainties. 
To do this, we made two additional computations but including a cool spot 
either on the primary or on the secondary and searching the entire surfaces for its location. 
It turned out that the results did not change much when we assumed the spot to be 
on the surface of the secondary star. However, the solution with a cool spot placed on the 
primary component resulted in the temperature of the secondary component being about
300~K smaller than that of the primary, while other parameters being similar to these obtained
in the non-spotted solution. As expected, the theoretical light curves of spotted solutions 
fit observed ones better, as the number of free parameters was larger.

\begin{table}
\caption{Absolute parameters of the contact system BO CVn} 
\center
\begin{tabular}{lcc}
\hline
%\textbf{Parameter} & \textbf{Primary} &  \textbf{Secondary} \\
Parameter & Primary &  Secondary \\
\hline
$A$ ($R_{\odot}$) & \multicolumn{2}{c}{3.14$\pm$0.05} \\
$M$ ($M_{\odot}$) & 1.16${\pm}$0.04    & 0.39${\pm}$0.02 \\
$R$ ($R_{\odot}$) & 1.62${\pm}$0.03    & 1.00${\pm}$0.02 \\
%$T$ (K)           & 6980${\pm}$222     & 7191${\pm}$222 \\
$L$ ($L_{\odot}$) & 5.53${\pm}$0.37    & 2.38${\pm}$0.09 \\
\hline
\end{tabular}
\label{TabAbsParam}
\end{table}

We analyzed the O-C diagram of BO CVn based on the literature data appended with a new time
of primary minimum determined from our observations. 
Although we found the period was not constant, we found no short-period behavior 
that would provide support for a companion to the binary system.
Thus third light may or may not represent a star gravitationally bound to the binary
system.
The O-C diagram shows an upward parabola that could be 
a result of period increase.
This could be interpreted as mass transfer between the components, in a conservative case
such a period increase would correspond to 6.3$\times$10$^{\rm -8}$ $M_{\odot}$/year. 
However, if the third star orbits the binary system in a very wide orbit, this might also  be
a part of a cyclic behavior due to the light travel effect.
 
We conclude that in several aspects, BO CVn is very similar to \astrobj{UZ Leo} 
\citep{zola10} - both systems have similar spectral types, mass ratios and high 
filling factors. 
Also for both, a  considerable third light was found and their 
O-C diagrams (upward parabolas) indicate periods increase, therefore excluding mass loss from
the system.

\section{Acknowledgments}
We acknowledge a helpful discussion with Bob Wilson. This research has made use of the 
SIMBAD database, operated at CDS, Strasbourg, France.

\bibliographystyle{elsarticle-harv}
\bibliography{refer}

\begin{thebibliography}{00}

%% \bibitem must have one of the following forms:
%%   \bibitem[Jones et al.(1990)]{key}...
%%   \bibitem[Jones et al.(1990)Jones, Baker, and Williams]{key}...
%%   \bibitem[Jones et al., 1990]{key}...
%%   \bibitem[\protect\citeauthoryear{Jones, Baker, and Williams}{Jones
%%       et al.}{1990}]{key}...
%%   \bibitem[\protect\citeauthoryear{Jones et al.}{1990}]{key}...

\bibitem[\protect\citeauthoryear{Agerer \& H{\"u}bscher}{1996}]{agerer96} Agerer, F., H{\"u}bscher, J., 1996, IBVS, 4382, 1
\bibitem[\protect\citeauthoryear{Agerer \& H{\"u}bscher}{1999}]{agerer99} Agerer, F., H{\"u}bscher, J., 1999, IBVS, 4711, 1
\bibitem[\protect\citeauthoryear{Agerer \& H{\"u}bscher}{2000}]{agerer00} Agerer, F., H{\"u}bscher, J., 2000, IBVS, 4912, 1
\bibitem[\protect\citeauthoryear{Agerer \& H{\"u}bscher}{2001}]{agerer01} Agerer, F., H{\"u}bscher, J., 2001, IBVS, 5016, 1
\bibitem[\protect\citeauthoryear{Albayrak et al.}{2000}]{albayrak00} Albayrak, B., M{\"u}yessero\u{g}lu, Z., {\"O}zdemir, S., 2000, IBVS, 4941, 1
\bibitem[\protect\citeauthoryear{Albayrak et al.}{2001}]{albayrak01} Albayrak, B., M{\"u}yessero\u{g}lu, Z., {\"O}zdemir, S., 2001, AN, 322, 125
\bibitem[\protect\citeauthoryear{Albayrak et al.}{2002}]{albayrak02} Albayrak, B., Tanr{\i}verdi, T., Ayd{\i}n C., 2002, IBVS, 5300, 1
\bibitem[\protect\citeauthoryear{Albayrak et al.}{2005}]{albayrak05} Albayrak, B., Y{\"u}ce, K., Selam, S.O., Tanr{\i}verdi, T., Okan, A., \c{C}{\i}nar, D., Topal, S., {\"O}zg{\"u}r, E., \c{S}ener, H.T., Erg{\"u}n, \.{I}., Civelek, E., 2005, IBVS, 5649, 1
\bibitem[\protect\citeauthoryear{Br{\'a}t et al.}{2007}]{brat07} {Br{\'a}t}, L., {Zejda}, M., {Svoboda}, P., 2007, OEJV, 74, 1
\bibitem[\protect\citeauthoryear{Br{\'a}t et al.}{2008}]{brat08} {Br{\'a}t}, L., {{\v S}melcer}, L, {Ku{\`e}{\'a}kov{\'a}}, H., {Ehrenberger}, R., {Koci{\'a}n}, R., {Lomoz}, F., {Urban{\`e}ok} L., {Svoboda}, P., {Trnka}, J., {Marek}, P., {D{\o}ev{\`i}n{\'y}}, R., {Uhl{\'a}{\o}}, R., {Poddan{\'y}}, S., {Zasche}, P., {Skarka}, M., 2008, OEJV, 94, 1
\bibitem[\protect\citeauthoryear{Br{\'a}t et al.}{2009}]{brat09} {Brat}, L., {Trnka}, J., {Lehky}, M., {Smelcer}, L., {Kucakova}, H., {Ehrenberger}, R., {Dreveny}, R., {Lomoz}, F., {Marek}, P., {Kocian}, R., {Svoboda}, P., {Pribik}, V., {Urbancok}, L., {Poddany}, S., {Dubovsky}, P.~A., {Uhlar}, R., {Horalek}, P., {Hanzl}, D., {Broz}, M., {Kalisch}, T., {Macek}, O., {Exnerova}, M., {Vitek}, M., 2009, OEJV, 107, 1
\bibitem[\protect\citeauthoryear{D\'{\i}az-Cordov\'{e}s et al.}{1995}]{diaz95} D\'{\i}az-Cordov\'{e}s, J., Claret, A., Gimenez, A., 1995, A\&AS, 110, 329
\bibitem[\protect\citeauthoryear{Harmanec}{1988}]{harmanec88} Harmanec, P., 1988, BAICz, 39, 329
\bibitem[\protect\citeauthoryear{H{\"u}bscher et al.}{1991}]{hubscher91} H\"{u}bscher, J., Agerer, F., Wunder, E., 1991, BAV Mitt. 59
\bibitem[\protect\citeauthoryear{H{\"u}bscher et al.}{1992}]{hubscher92} H\"{u}bscher, J., Agerer, F., Wunder, E., 1992, BAV Mitt. 60
\bibitem[\protect\citeauthoryear{H{\"u}bscher et al.}{1993}]{hubscher93} H\"{u}bscher, J., Agerer, F., Wunder, E., 1993, BAV Mitt. 62
\bibitem[\protect\citeauthoryear{H{\"u}bscher et al.}{1994}]{hubscher94} H\"{u}bscher, J., Agerer, F., Frank, P., Wunder, E., 1994, BAV Mitt. 68
\bibitem[\protect\citeauthoryear{H{\"u}bscher et al.}{2005}]{hubscher05} H\"{u}bscher, J., Paschke, A., Walter, F., 2005, IBVS, 5657, 1
\bibitem[\protect\citeauthoryear{H{\"u}bscher \& Monninger}{2011}]{hubscher11} H\"{u}bscher, J., Monninger, G., 2011, IBVS, 5959, 1
\bibitem[\protect\citeauthoryear{Kreiner}{2004}]{kreiner04} Kreiner, J.M., 2004, AcA, 54, 207
\bibitem[\protect\citeauthoryear{M{\"u}yessero\u{g}lu et al.}{2003}]{muyes03} M{\"u}yessero\u{g}lu, Z., T{\"o}r{\"u}n, E., {\"O}zdemir, T., G{\"u}rol, B., {\"O}zavc{\i}, \.{I}., Tun\c{c}, T., Kaya, F., 2003, IBVS, 5463, 1
\bibitem[\protect\citeauthoryear{Nagai}{2005}]{nagai05} Nagai, K., 2005, VSOLJ, 43 
\bibitem[\protect\citeauthoryear{Nagai}{2009}]{nagai09} Nagai, K., 2009, VSOLJ, 48
\bibitem[\protect\citeauthoryear{Nelson et al.}{2006}]{nelson06} Nelson, R.H. ,Terrell, D., Gross, J., 2006, IBVS, 5715, 1
\bibitem[\protect\citeauthoryear{Nelson}{2008}]{nelson08} Nelson, R.H., 2008, IBVS, 5820, 1
\bibitem[\protect\citeauthoryear{Nelson}{2009}]{nelson09} Nelson, R.H., 2009, IBVS, 5875, 1
\bibitem[\protect\citeauthoryear{Nelson}{2010a}]{nelson10a} Nelson, R.H., 2010a, IBVS, 5929, 1
\bibitem[\protect\citeauthoryear{Nelson}{2010b}]{nelson10b}Nelson, R.H., 2010b, "Spectroscopy for Eclipsing 
Binary Analysis in The Alt-Az Initiative, Telescope Mirror \& Instrument Developments", 
in: Genet, J.M. Johnson and V. Wallen (Eds.), Collins Foundation Press, Santa Margarita, CA  
\bibitem[\protect\citeauthoryear{Oja}{1989}]{oja89} Oja, T., 1989, IBVS, 3288, 1
\bibitem[\protect\citeauthoryear{Qian \& Zhu}{2006}]{qian06} Qian, S.-B., Zhu, L.-Y., 2006, AJ, 131, 1032
\bibitem[\protect\citeauthoryear{Rucinski}{2004}]{rucinski04} Rucinski, S.M., 2004, "Advantages of the Broadening Function (BF) over the Cross-Correlation Function (CCF)", in Stellar Rotation, Proc. IAU Symp. 215
\bibitem[\protect\citeauthoryear{Selam et al.}{2003}]{selam03} Selam, S.O., Albayrak, B., \c{S}enavc{\i}, H.V., Tanr{\i}verdi, T., Elmasl{\i}, A., Kara, A., Aksu, O., Y{\i}lmaz, M., Karaka\c{s}, T., \c{C}{\i}nar, D., Demirhan, M., \c{S}ahin, S., \c{C}eviker, S., G{\"o}zler, A.P., 2003,  IBVS, 5471, 1
\bibitem[\protect\citeauthoryear{Strassmeier et al.}{2010}]{strassmeier10} Strassmeier, K.G., Granzer, T., Kopf, M., Weber, M., K{\"u}ker, M., Reegen, P., Rice, J.~B., Matthews, J.~M., Kuschnig, R., Rowe, J.~F., Guenther, D.~B., Moffat, A.~F.~J., Rucinski, S.~M., Sasselov, D., Weiss, W.~W., 2010, A\&A, 520, 52
\bibitem[\protect\citeauthoryear{Terrell \& Wilson}{2005}]{terrellwilson2005} Terrell, D. Wilson, R.E., 2005, Ap\&SS, 296, 221 
\bibitem[\protect\citeauthoryear{Wilson \& Devinney}{1971}]{wilson71} Wilson, R.E., Devinney, E.J., 1971, ApJ, 166, 605 
\bibitem[\protect\citeauthoryear{Wilson}{1979}]{wilson79} Wilson, R.E., 1979, ApJ, 234, 1054
\bibitem[\protect\citeauthoryear{Wilson}{1993}]{wilson93} Wilson, R.E., 1993, Documentation of Eclipsing Binary Computer Model, University of Florida
\bibitem[\protect\citeauthoryear{Y{\i}lmaz et al.}{2009}]{yilmaz09} Y{\i}lmaz, M., Ba\c{s}turk, {\"O}., 
Alan, N., \c{S}enavc{\i}, H.V., Tanr{\i}verd{\i}, T., K{\i}l{\i}\c{c}o\u{g}lu, T., \c{C}al{\i}\c{s}kan, S., 
\c{C}elik, L., Ayd{\i}n, G., \c{C}akan, D., Bilgi\c{c}, D., Ulu\c{s}, N.D., Elmasl{\i}, A., 
Selam, S.O., Albayrak, B., Ekmek\c{c}i, F., 2009, IBVS, 5887, 1
\bibitem[\protect\citeauthoryear{Zola et al.}{2010}]{zola10} Zola, S., Gazeas, K., Kreiner, J.M., 
Ogloza, W., Siwak, M., Koziel-Wierzbowska, D., Winiarski, M.,  2010, MNRAS, 408, 464 
%%


\end{thebibliography}



\end{document}